\documentstyle[epsfig, 12pt,preprint,tighten,aps]{revtex}

\begin{document}
\begin{titlepage}

\rightline{{\large \tt November 2002}}
\rightline{{\large \tt astro-ph/0211067}}

\vskip 1.4 cm

\centerline{\Large \bf
Mirror matter in the Solar system:}
\centerline{\Large \bf
New evidence for mirror matter from Eros} 

\vskip 1.2 cm

\centerline{\large R. Foot$^a$ and S. Mitra$^b$}
\vskip 0.7 cm\noindent
\centerline{{\large $^a$\ \it foot@physics.unimelb.edu.au}}
\centerline{{\large \it School of Physics}}
\centerline{{\large \it Research Centre for High Energy Physics}}
\centerline{{\large \it The University of Melbourne}}
\centerline{{\large \it Victoria 3010 Australia }}
\vskip 0.3cm
\centerline{{\large $^b$\ \it saibalm@science.uva.nl}}
\centerline{{\large \it Instituut voor Theoretische Fysica}}
\centerline{{\large \it Universiteit van Amsterdam}}
\centerline{{\large \it 1018 XE Amsterdam}}
\centerline{{\large \it The Netherlands}} 

\vskip 1.4cm

\centerline{\large Abstract} \vskip 0.5cm \noindent
Mirror matter is an entirely new form of matter predicted
to exist if mirror symmetry is a fundamental symmetry
of nature. Mirror matter has the right broad properties to
explain the inferred dark matter of the Universe and might
also be responsible for a variety of other puzzles in 
particle physics, astrophysics, meteoritics and planetary science.
It is known that mirror matter can interact with ordinary matter
non-gravitationally via photon-mirror photon kinetic mixing.
The strength of this possibly fundamental interaction depends on the
(theoretically) free parameter $\epsilon$.
We consider various proposed manifestations of mirror
matter in our solar system examining in particular how the
physics changes for different possible values of $\epsilon$. 
We find new evidence
for mirror matter in the solar system coming from the observed sharp
reduction in crater rates (for
craters less than about 100 meters in diameter)
on the asteroid 433 Eros.
We also re-examine various existing ideas including
the mirror matter explanation for the anomalous meteorite
events, anomalous slow-down of Pioneer spacecraft etc.  

\end{titlepage}


\section{Introduction}

One of the most obvious candidates for a fundamental symmetry of nature is
mirror symmetry. While
it is an established experimental fact that mirror symmetry
appears broken by the interactions of the known elementary
particles (because of their left-handed weak interactions), 
this however does not exclude the possible existence of
exact unbroken mirror symmetry in nature. This is because mirror
symmetry (and also time reversal) can be exactly conserved if a set of
mirror particles exist\cite{ly,flv}.  The idea is that for each
ordinary particle, such as the photon, electron, proton and
neutron, there is a corresponding mirror particle, of exactly the
same mass as the ordinary particle. 
Furthermore, the mirror particles interact with each other in
exactly the same way that the ordinary particles do. 
The mirror particles are not produced
(significantly) in laboratory experiments just because they couple
very weakly to the ordinary particles. In the modern language of
gauge theories, the mirror particles are all singlets under the
standard $G \equiv SU(3)\otimes SU(2)_L \otimes U(1)_Y$ gauge
interactions. Instead the mirror fermions interact with a set of
mirror gauge particles, so that the gauge symmetry of the theory
is doubled, i.e. $G \otimes G$ (the ordinary particles are, of
course, singlets under the mirror gauge symmetry)\cite{flv}.
Mirror symmetry is conserved because the mirror fermions experience $V+A$
(right-handed) mirror weak interactions and the ordinary fermions
experience the usual $V-A$ (left-handed) weak interactions.
Ordinary and mirror particles interact with each other
predominately by gravity only.

At the present time there is an interesting range of experimental
observations supporting the existence of mirror matter, for a
review see Ref.\cite{comet,puz} (for a more detailed discussion
of the case for mirror matter, accessible also to
the non-specialist, see the recent book\cite{bk}).  
The evidence includes numerous observations
suggesting the existence of invisible `dark matter' in galaxies.
Mirror matter is stable and dark and provides a natural candidate
for this inferred dark matter\cite{blin}. 
The MACHO observations\cite{mo}, close-in extrasolar
planets\cite{ce}, isolated planets\cite{is}, 
gamma ray bursts\cite{grb} and even the comets\cite{comet,bk} may
all be mirror world manifestations.

While we know that ordinary and mirror
matter do not interact with each other via any
of the {\it known} non-gravitational forces,
it is possible that new interactions exist which
couple the two sectors together.
In Ref.\cite{flv,flv2}, all such interactions consistent
with gauge invariance, mirror symmetry and renormalizability
were identified, namely,
photon-mirror photon kinetic mixing, Higgs-mirror Higgs
interactions and via ordinary neutrino-mirror neutrino
mass mixing (if neutrinos have mass).
Of most importance though for this paper is
the photon-mirror photon kinetic mixing interaction.

In quantum field theory,
photon-mirror photon kinetic mixing
is described by the interaction
\begin{equation}
{\cal L} = {\epsilon \over 2}F^{\mu \nu} F'_{\mu \nu},
\label{ek}
\end{equation}
where $F^{\mu \nu}$ ($F'_{\mu \nu}$) is the field strength tensor
for electromagnetism (mirror electromagnetism). This type of
Lagrangian term is gauge invariant and renormalizable and can
exist at tree level\cite{flv,fh} or may be induced radiatively in
models without $U(1)$ gauge symmetries (such as grand unified
theories)\cite{gl,bob,cf}. One effect of ordinary photon-mirror
photon kinetic mixing is to give the mirror charged particles a
small electric charge\cite{flv,gl,bob}. That is, they couple to
ordinary photons with electric charge $\epsilon e$.

The most important 
experimental particle physics implication of 
photon-mirror
photon kinetic mixing is that it modifies the properties of 
orthopositronium\cite{gl}. The current experimental
situation is summarized in Ref.\cite{fg}, which
shows that $\epsilon \stackrel{<}{\sim} 10^{-6}$,
with some evidence for $\epsilon \approx 10^{-6}$ from
the 1990 vacuum cavity experiment\cite{vac}.
Photon-mirror photon kinetic mixing has other important
implications, including astrophysical effects\cite{oimp},
the formation
of ordinary atom-mirror atom bound states\cite{oimp2} and
implications for early Universe cosmology\cite{cg}. 

If there is some amount of mirror matter in the solar
system then the photon-mirror photon kinetic mixing force
will cause unique effects when mirror matter space-bodies (SB) collide
with the Earth\cite{tunguska,fy} and also when ordinary spacecraft propagate
out through the solar system\cite{fvpioneer}. 
Motivation to study such effects
comes from observations of anomalous meteorite events on both
small\cite{docobo,jas} and large scales\cite{val}, as well as
the observed anomalous slow-down of {\it both} Pioneer spacecraft\cite{study}.
While previous work focussed on the case of relatively large
$\epsilon \sim 10^{-6}$, the aim of the present paper
is to study the implications more generally; exploring how
things change for smaller values of $\epsilon$.
We also consider the effects of mirror SB impacts on ordinary asteroids.
It is pointed out that collisions between asteroids and small mirror matter SB below a certain 
critical size, which
we estimate as a function of $\epsilon$, will not lead to the formation
of an impact crater. Thus, we are able to predict the existence of
a crater hiatus. Such a crater reduction rate has in fact been
observed by spacecraft studies on Eros which indicates a value
of $\epsilon \sim 10^{-7}-10^{-9}$ if this mirror matter
interpretation is correct. Such small values of $\epsilon$ are 
consistent with the observations of anomalous Earth impact events, Pioneer
spacecraft anomaly and
is also potentially testable in future orthopositronium experiments
(such as the ETH-Moscow orthopositronium experiment\cite{eth}).

\section{Impacting mirror Space-body with the 
Earth: Theory}

There is not much room for a large amount of mirror matter in our
solar system. For example, the amount of mirror matter within the
Earth has been constrained to be less than $10^{-3}
M_{Earth}$\cite{sashaV}. However, we don't know enough about the
formation of the solar system to be able to exclude the existence
of a large number of  Space Bodies (SB) made of mirror matter if
they are small like comets and asteroids. The total mass of
asteroids in the asteroid belt is estimated to be only about
0.05\% of the mass of the Earth. A similar or even greater number
of mirror bodies, perhaps orbiting in a different plane or even
spherically distributed like the Oort cloud is a fascinating 
possibility\footnote{ Large planetary sized
bodies are also possible if they are in distant
orbits\cite{silnem} or masquerade as ordinary planets or moons
by accreting ordinary matter onto
their surfaces\cite{bk}.}. 

An even more remarkable possibility is that the comets themselves
are {\it the}
mirror matter SB\cite{bk,comet}. Of course, comets must contain at least some
ordinary matter component (which could easily have been
accreted over time) to explain the coma/tail phenomena, however
observations are consistent with a small ordinary matter
component which typically gets depleted after the comet enters
the inner solar system for the first time.
In this way the long standing comet fading problem can be solved: Many
comets lose their small ordinary matter component after
passing into the inner solar system leaving an
almost invisible mirror matter core. Furthermore, this 
idea is consistent with recent observations
confirming the apparent absence of any visible cometary
remnants\cite{lev}.

Anyway, if small mirror matter bodies do exist and happen to collide
with the Earth, what would be the consequences?
If the only force connecting mirror matter
with ordinary matter is gravity, then the consequences
would be minimal. The mirror matter space-body would simply pass
through the Earth and nobody would know about it unless
the body was so heavy as to gravitationally affect the motion
of the Earth. However, it is possible for ordinary and mirror
particles to interact with each other by new interactions.
Of most importance (for macroscopic bodies) is the
photon-mirror photon transition force, Eq.(\ref{ek}). The effect of
this interaction is to make mirror charged particles
(such as mirror electron and mirror proton) couple
to the ordinary photon with effective electric charge
$\epsilon e$\cite{flv,gl,bob}.

One effect of this interaction is that the mirror nuclei 
can interact with the ordinary nuclei, as illustrated below:
\vskip 0.5cm
\centerline{\epsfig{file=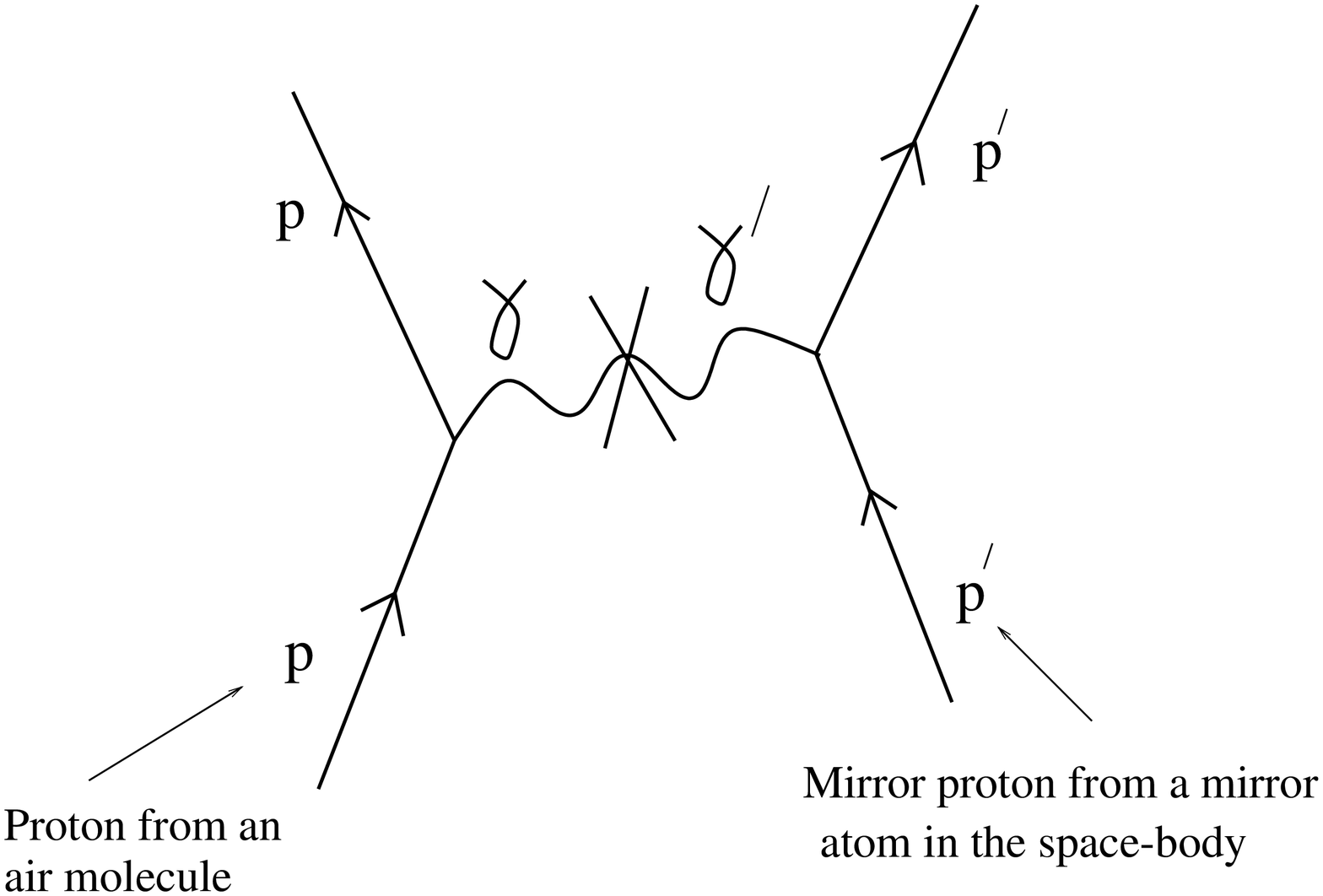,width=9.7cm}}
\vskip 0.5cm
\noindent 
{\small Figure 1: Rutherford elastic scattering of air molecules
as they pass through the mirror matter space-body.  The X represents
the photon-mirror photon kinetic mixing interaction.}
\vskip 1.0cm
\noindent
In other words, the nuclei of the mirror atoms of the space-body 
will undergo Rutherford scattering with the nuclei of the 
atmospheric nitrogen and oxygen atoms. In addition, ionizing 
interactions can occur (where electrons are removed from the atoms)
which can ionize both the mirror atoms and also
the (ordinary) atmospheric atoms.  
This would make the mirror matter space-body effectively visible 
as it plummets to the surface of our planet. 

The effect of the atmosphere upon the mirror
SB depends on a number
of factors, including, the strength of the photon-mirror
photon transition force ($\epsilon$), the chemical
composition of the space-body, its initial velocity
and its size and shape.
We could estimate the initial velocity of the
space-body by observing that the velocity of the Earth
around the Sun is about 30 km/s. The space-body should
have a similar velocity so that depending on its direction,
the relative velocity of the space-body when viewed from
Earth would be expected to be between about 11 and 70 km/s
\footnote{
The minimum velocity of a space-body as
viewed from Earth is not zero because of the
effect of the local gravity of the Earth. 
The minimum velocity of a space-body
is the Earth's escape velocity, about 11 km/s.}.

Previous work\cite{tunguska} has shown that
for relatively large values of $\epsilon \sim 10^{-6}$,
a mirror matter SB impacting with the atmosphere
will experience the standard drag force:
\begin{eqnarray} 
F_{drag} = C_d \rho_{air} S v^2/2
\label{st}
\end{eqnarray}
where $\rho_{air}$ is the mass density of the air, $v$ is the
velocity of the SB relative to the Earth and $S$ is the cross sectional area.
The reason is that the nuclei of the air molecules Rutherford scatter
with the mirror nuclei
of the mirror atoms which make up the mirror SB. Provided
that the scattering length is much less than
the dimensions of the SB -- which happens to
be the case for large $\epsilon \sim 10^{-6}$ -- the momentum transferred
and hence the drag force on the mirror SB will be roughly
the same as the standard case of an ordinary matter SB.
[In Eq.(\ref{st}), $C_d$ is the drag coefficient, which
incorporates the hydrodynamic
properties of air flow around the body. It depends on the
shape, velocity and other factors, but is typically
of order 1].

Obviously if $\epsilon$ is small enough, the interactions
of the air molecules with the SB will become weak enough so that
the air molecules can pass through the SB without losing much
of their momentum (in the rest frame of the SB). In this
regime, the drag force is proportional to the total number
of molecules within the SB rather than the bodies cross 
sectional surface area ($S$).
Our purpose here is to study in detail the
effect of the atmosphere on the SB as a function
of the (fundamental) parameter $\epsilon$, and the other
variables such as velocity, SB size etc. 
This section can be viewed as a generalization
of the previous 
paper\cite{tunguska}, which focussed mainly on the case of
large $\epsilon \sim 10^{-6}$.

Assume that the mirror matter SB is composed of atoms
of mass $M_{A'}$ and the air is composed of molecules 
of mass $M_A$. The (mirror) electric charge in units
of $e$ of the (mirror) nuclei, 
will be denoted as $Z$ ($Z'$). Let us assume that the 
trajectory of the SB is a straight line along the $\hat{z}$
axis of our co-ordinate system. In the rest frame of the SB,
the change in forward momentum of each of the on-coming
atmospheric molecules is then
\begin{eqnarray}
{dP_z \over dt} = \Gamma_{coll} M_A (v\cos\theta_{scatt} - v) =
-2\Gamma_{coll}M_A v \sin^2 {\theta_{scatt} \over 2}
\label{rr}
\end{eqnarray}
where $\theta_{scatt}$ is the scattering angle in 
the rest frame of the SB and $\Gamma_{coll}$ is the
collision rate of the atmospheric molecules with the mirror
atoms of the SB. 
The collisions also generate transverse
momentum, which we can approximately neglect for the moment since it
averages out to zero (which means that we can replace $v$
by $v_z$ below). 

The collision rate, $\Gamma_{coll}$ can
be related to the cross section in the usual way,
$\Gamma_{coll} = \sigma v_z \rho_{SB}/M_{A'}$, and thus
Eq.(\ref{rr}) becomes
\begin{eqnarray}
{dP_z \over dt} = -2\left( {M_A \over M_{A'}} \right)
\int {d\sigma \over d\Omega} \rho_{SB} v^2_z \sin^2 {\theta_{scatt} \over 2}
d\Omega
\label{rr2}
\end{eqnarray}
There are two main processes which can contribute to the 
scattering cross section. For the velocities of
interest, $v \stackrel{<}{\sim} 70$ km/s, the
cross section is dominated by Rutherford scattering
of the mirror nuclei of effective electric charge
$\epsilon Z' e$ with the ordinary nuclei of
electric charge $Ze$, modified for small angle
scattering by the screening effects of the atomic
electrons (at roughly the radius $r_0 \approx 10^{-8}-10^{-10}
$ cm, depending on atomic number). It is given by
\footnote{This equation includes
a factor of 2, because each atmospheric $N_2$ or $O_2$ molecule has 
2 nuclei. Also, natural units with $\hbar = c = 1$ are used unless
otherwise stated.}\cite{mer}:
\begin{eqnarray}
{d \sigma \over d\Omega} = {8 M_A^2 \epsilon^2 e^4 Z^2 Z'^2 
\over (4M_A^2 v_z^2 \sin^2 {\theta_{scatt}\over 2} + {1 \over
r_0^2})^2}.
\label{rr3}
\end{eqnarray}
Thus, combining Eq.(\ref{rr2}) and Eq.(\ref{rr3}), we
end up with the following differential equation
describing the motion of an air molecule
within the SB:
\begin{eqnarray}
{dP_z \over dt} = M_A v_z {dv_z \over dz} \simeq -{8\pi Z^2 Z'^2 \rho_{SB}
\epsilon^2 e^4 \over M_{A'} M_A v_z^2}
log_e (M_A v_z r_0),
\end{eqnarray}
which is valid for $M_A v r_0 \gg 1$.
For $M_A \approx 30 M_P$ [$M_P$ is the proton mass],
$M_A v r_0 \approx 150 (v/30 {\rm km/s})(r_0/10^{-9} {\rm cm})$
which means that the above equation is generally valid for
the velocities of interest.
Replacing the log factor by a number of order $log_e (150) \sim 5$,
the above equation can be easily solved to give the
velocity ($v_f$) of an air molecule after travelling a distance
$z$ within the body:
\begin{eqnarray}
{v_i^4 \over 4} - {v_f^4 \over 4} =
{Z^2 Z'^2 \rho_{SB} \epsilon^2 e^4 40\pi z\over M_{A'} M_{A}^2}.
\label{mon}
\end{eqnarray}
Note that $v_i$ is the initial velocity of the air molecule with
respect to the SB [which is obviously equal in magnitude to the velocity of
the SB when viewed from Earth].
The condition that the air molecules are effectively `stopped'
within the SB is that $v_f \approx 0$\footnote{
Of course, the air molecules will not be completely stopped,
but will end up with a thermal velocity
of $v \sim \sqrt{kT/M_A} \sim 600 \sqrt{T/1000 K} \ {\rm m/s}$.},
which gives a distance
\begin{eqnarray}
z \approx {v^4 M_A^2 M_{A'} \over 160\pi Z^2 Z'^2 \rho_{SB} \epsilon^2 e^4}
\approx 3\left( {10^{-6} \over \epsilon}\right)^2 \left( {v \over
30 \ {\rm km/s}}\right)^4 \ {\rm millimetres}. 
\label{stop}
\end{eqnarray}
The right-hand side of the above equation is roughly valid for a SB made of 
mirror ice. For heavy elements, such as mirror iron or nickel the right-hand side is about 10
times smaller.

Obviously, the air molecules can only be stopped within
the SB provided that the stopping distance ($z$) is
less than the diameter of the SB ($D_{SB}$).
Thus, there are essentially two regimes of interest,
which we will call the {\it strong coupling
regime} ($D_{SB}/z \stackrel{>}{\sim} 1$) and {\it weak coupling regime}
($D_{SB}/z \stackrel{<}{\sim} 1$).

\subsection{The strong coupling regime}

In the strong coupling regime, 
the air molecules lose a significant fraction of their
momentum after penetrating the SB, which implies
an atmospheric drag force of the standard form, Eq.(\ref{st}).
The condition for this to occur is that
$D_{SB}/z \stackrel{>}{\sim} 1$, that is,
\begin{eqnarray}
{D_{SB}160\pi Z^2 Z'^2 \rho_{SB} \epsilon^2 e^4 \over
v^4 M_A^2 M_{A'} }
\approx \left( {D_{SB} \over 5\ {\rm metres}}\right)\left( {\epsilon \over
2\times 10^{-8}}\right)^2 \left( {30\ {\rm km/s} \over v}\right)^4
\stackrel{>}{\sim} 1
\nonumber \\
.[Strong \ coupling \ regime].
\label{y76}
\end{eqnarray}
Note that because $v$ can only decrease, a SB 
initially in the strong coupling regime will always
remain in the strong coupling regime.

In this strong coupling regime the drag force has the
(standard) form Eq.(\ref{st}),
which can easily be shown to lead to an
exponentially decaying SB velocity:
\begin{eqnarray}
v_{SB} &=& v_{SB}^i e^{\left(-{x\over L}\right)},
\nonumber \\
&=& v_{SB}^i  e^{\left( -{h \over H}\right)}
\end{eqnarray}
where $h=x\cos\theta$, $H = L\cos\theta$ ($\theta$ is
the zenith angle of the SB, as illustrated in
Figure 2) and
\begin{eqnarray}
H = {x \cos\theta \over \int^x {\rho_{atm} S C_d dx\over 2M_{SB} }}.
\end{eqnarray}
\vskip 0.7cm
\centerline{\epsfig{file=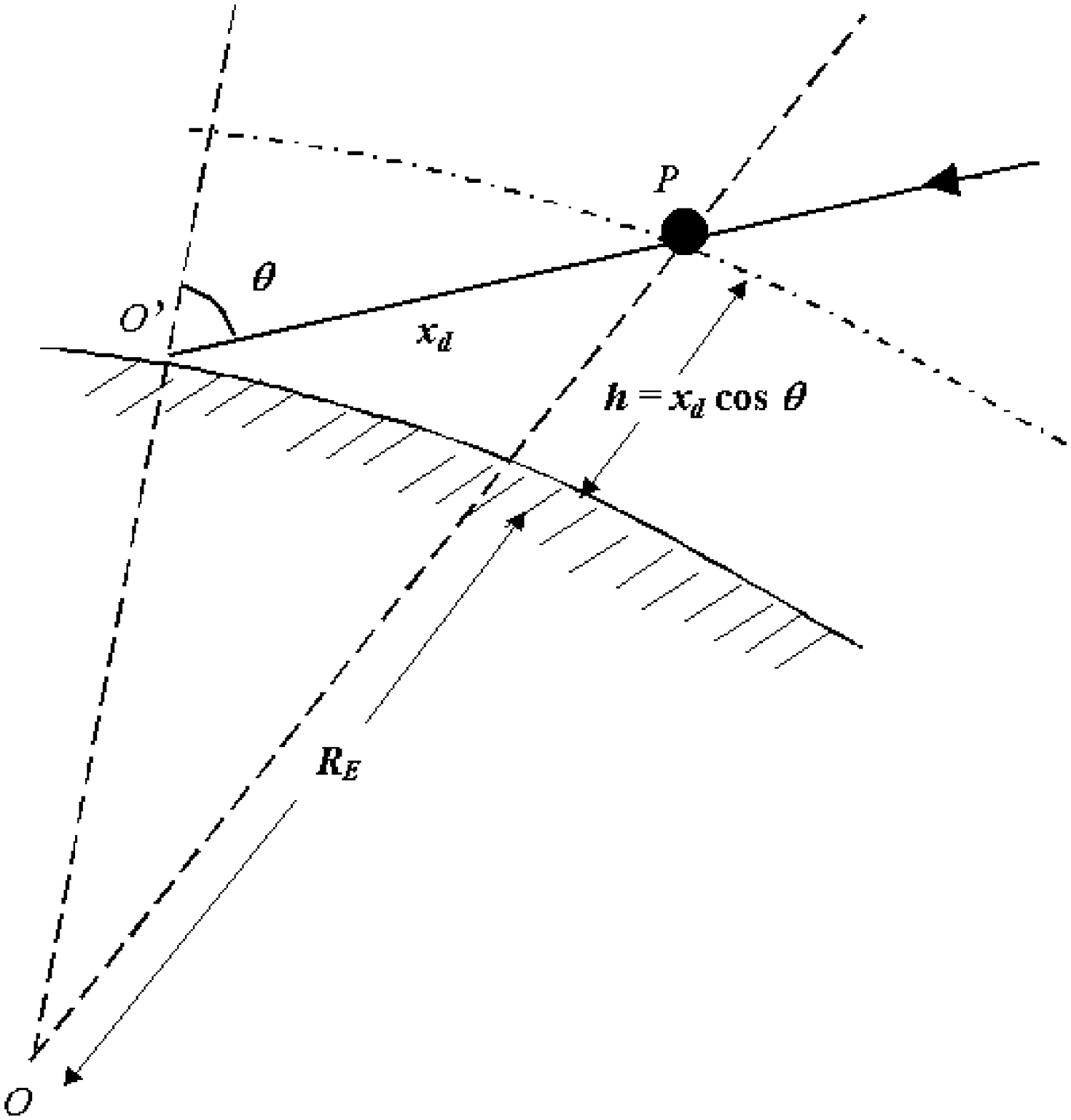,height=7cm,width=8cm}}
\vskip 0.5cm
\noindent 
{\small 
Figure 2: Trajectory of a SB entering the Earth's atmosphere,
taken to be approximately a straight path.} 
\vskip 1cm
\noindent
For the special case of constant air density (and $C_d \sim 1$), 
$H = 2\cos\theta D_{SB}\rho_{SB}/\rho_{atm}$, where $D_{SB} \equiv V/S$
is a measure of the size of the SB.
Of course, the air density is exponentially decreasing with a scale
height of $h_0 \approx 8$ km [i.e. $\rho_{atm} =
\rho^0_{atm} e^{-h/h_0},\ \rho_{atm}^0 \simeq 1.2 \times 10^{-3}\
{\rm g/cm^3}$ is the air density at sea-level].
Thus, we see that there is a critical SB size
[$\stackrel{\sim}{D}_{SB}(\theta)$]
given approximately by
\begin{eqnarray}
\stackrel{\sim}{D}_{SB}(\theta) \approx {h_0 \rho_{atm}^0 \over 2\cos\theta \rho_{SB} }
\sim {5 \ {\rm metres} \over \cos\theta} .
\label{pp2}
\end{eqnarray}
Space-bodies smaller than about $\stackrel{\sim}{D}_{SB}$ lose
their initial velocity in the atmosphere, while space-bodies
larger than $\stackrel{\sim}{D}_{SB}$ retain most of their
velocity.
In other words, in the strong coupling regime there are two limiting cases
of interest:
\begin{eqnarray}
(a) \ \ &D_{SB}& \stackrel{<}{\sim} \ \stackrel{\sim}{D}_{SB} \ 
\sim {5 \ {\rm metres} \over \cos\theta} 
\Rightarrow v^f_{SB} \ll v_{SB}^i
\nonumber \\
(b) \ \ &D_{SB}& \stackrel{>}{\sim} \ \stackrel{\sim}{D}_{SB} \ \sim {5 \ {\rm metres} 
\over \cos\theta}
\Rightarrow v^f_{SB} \sim v_{SB}^i
\label{df}
\end{eqnarray}
Note that the above conclusion applies also to an ordinary
matter SB. Of course, in practice there are complications due
to surface melting (ablation)  
and potential break up of the body (which we will return to
in the following section).

Before we investigate what happens in the weak coupling regime,
let us first introduce the following quantity: 
\begin{eqnarray}
\omega &\equiv &
{\stackrel{\sim}{D}_{SB} (\theta = 0) \over z} =
{ 160\pi Z^2 Z'^2 \rho_{SB} \epsilon^2 e^4 \stackrel{\sim}{D}_{SB}(\theta = 0) \over
M_A^2 M_{A'} v^4} 
\nonumber \\
& = &
{80\pi Z^2 Z'^2 \rho_{atm}^0 h_0 \epsilon^2 e^4 \over
M_A^2 M_{A'} v^4}
\nonumber \\
& \sim &
\left({\epsilon \over 2\times 10^{-8}}\right)^2\left( {30\
{\rm km/s} \over v}\right)^4.
\end{eqnarray}
With this definition, the strong coupling regime
corresponds to
\begin{eqnarray}
\omega \stackrel{>}{\sim} {\stackrel{\sim}{D}_{SB}(\theta=0) \over D_{SB} } 
\approx {5\ {\rm metres} \over D_{SB}}
\ \ \ \ [Strong\ coupling \ regime]
\end{eqnarray}

\subsection{The weak coupling regime}

In the weak coupling regime, the air molecules
will pass through the SB without losing much of their
momentum. In this case the atmospheric drag force will
have quite a different form to Eq.(\ref{st}) (and we
will calculate it in a moment).
The weak coupling regime occurs when $D_{SB}/z \stackrel{<}{\sim} 1$,
that is,
\begin{eqnarray}
{D_{SB}160\pi Z^2 Z'^2 \rho_{SB} \epsilon^2 e^4 \over
v^4 M_A^2 M_{A'} }
\approx \left( {D_{SB} \over 5\ {\rm metres}}\right)\left( {\epsilon \over
2\times 10^{-8}}\right)^2 \left( {30 \ {\rm km/s} \over v}\right)^4
\stackrel{<}{\sim} 1
\nonumber \\
.[Weak \ coupling \ regime].
\nonumber \\
\label{y6}
\end{eqnarray}
or in terms of $\omega$,
\begin{eqnarray}
\omega \stackrel{<}{\sim} {5 \ {\rm metres} \over D_{SB} } 
\ \ \ \ [Weak \ coupling \ regime] 
\label{wc}
\end{eqnarray}
Note that if the velocity of the SB decreases significantly
then it is possible for the regime to change from the
weak coupling one to the strong coupling one. In fact, this
will happen if the SB loses its cosmic velocity in the atmosphere.

In the weak coupling regime, the air molecules cannot be stopped
within the SB, and hence do not transfer all of their momentum. 
The amount of momentum that each air molecule transfers 
(working again in the instantaneous rest frame of the SB)
is $M_A (v_i - v_f)$, which, in the 
weak coupling limit $D_{SB}/z \ll 1$, 
can be evaluated from Eq.(\ref{mon}):
\begin{eqnarray}
M_A (v_i - v_f) = {Z^2 Z'^2 \rho_{SB} \epsilon^2 e^4 40 \pi D_{SB} \over
M_{A'} M_A v_i^3}.
\label{ab}
\end{eqnarray}
The above equation tells us the amount of momentum lost
by the space body due to the impact of a single air molecule.
After travelling a distance $dx$, the SB will 
encounter $n(h) S dx$
air molecules (where $n(h) = \rho_{atm}/M_A$ is the number density of
air molecules at height $h$ above the ground). The 
net effect is a reduction in the SB forward
momentum of:
\begin{eqnarray}
dP_{SB} = M_A (v_i - v_f) n(h) S dx.
\end{eqnarray}
Using $P_{SB} = M_{SB}v_{SB}$
and Eq.(\ref{ab}),
the above equation can be re-written:
\begin{eqnarray}
{dv_{SB} \over dx} =
{Z^2 Z'^2 \epsilon^2 e^4 40\pi n(h) \over M_A M_{A'} v_{SB}^3
},
\label{yd}
\end{eqnarray}
where we have used $M_{SB} = \rho_{SB} S D_{SB}$
\footnote{
Technically, $M_{SB} = \rho_{SB}SD_{SB}$
is true only for a regular shape, however Eq.(\ref{yd}) actually 
holds more generally, using the fact that an irregular shape can be
written as a sum of regular shapes.}.
Note that Eq.(\ref{yd}) is independent of
the size parameters ($S, D_{SB}$)
which in the weak coupling regime can easily be understood
since each atom of the SB has an equal chance of interacting
with the on coming air molecules (which is, of course, in contrast
to the strong coupling regime).

Thus, in the weak coupling regime [defined by Eq.(\ref{y6})],
the equation governing the SB velocity, Eq.(\ref{yd}),
is independent of the size and shape of the body. Of course,
this is only strickly valid provided that the interactions
are weak enough and/or $n(h)$ is low
enough so that air pressure does not build up in front
of the SB.

To solve Eq.(\ref{yd}) we need to model the air number density
which can be simply done as follows. 
We assume that the atmosphere is composed
of molecules of mass $M_A \approx 30M_P$ ($M_P$ is the proton mass),
with number density profile:
\begin{eqnarray}
n(h) = n_0 exp\left( {-h \over h_0}\right),
\label{uu}
\end{eqnarray}
where $\rho_{atm}^0 \equiv M_A n_0 \simeq 1.2 \times 10^{-3}\ 
{\rm g/cm^3}$ is the
air mass density at sea-level, and $h_0 \approx 8$ km is the scale
height. Eq.(\ref{uu}), which can be derived from
hydrostatic equilibrium, is approximately
valid for $h \stackrel{<}{\sim} 25$ km. Above that
height, the density falls off more rapidly than given by
Eq.(\ref{uu}), but we will nevertheless use this
equation since it is a good enough approximation for the things
which we calculate.

We can now integrate Eq.(\ref{yd}) to obtain
$v_{SB}$ as a function of distance travelled
through the atmosphere. There are 
two limiting cases. First, the SB does not
lose significant velocity in the atmosphere.
In this case, $v_{SB}(h=\infty)-v_{SB}(h=0) \ll v_{SB}(h=\infty)$. 
This corresponds to:
\begin{eqnarray}
{Z^2 Z'^2 \epsilon^2 e^4 80\pi n_0 h_0 \over \cos\theta
M_{A} M_{A'} v_{SB}^4} &\ll & 1 
\nonumber \\
\Rightarrow  
\epsilon \stackrel{<}{\sim} 2\times 10^{-8} &\sqrt{cos\theta} &
\left( {v_{SB} \over 30 \ {\rm km/s}}
\right)^2 .
\label{qq}
\end{eqnarray}
Or, interms of $\omega$,
\begin{eqnarray}
\omega \stackrel{<}{\sim} \cos\theta \
\ \ \ [SB \ Retains \ velocity] 
\end{eqnarray}
On the other hand, the SB will lose most if its initial
velocity in the atmosphere, in the weak coupling regime,
if
\begin{eqnarray}
\omega \stackrel{>}{\sim} \cos\theta \
\ \ \ [SB\ Loses \ velocity] 
\label{lv}
\end{eqnarray}
Of course the system will pass from the weak coupling to
the strong coupling regime, as the SB loses velocity,
but this will not affect the conclusion that the SB
loses most of its initial velocity (one way to see
this is to note that Eq.(\ref{lv}) and Eq.(\ref{wc})
together imply that $D_{SB} \stackrel{<}{\sim} 5 \ {\rm
metres}/\cos\theta$. Recall, that this
is precisely the condition [Eq.(\ref{df})] that the SB loses its
velocity if it is in the strong coupling regime). 

To summarize things, we have identified four possible regimes
depending (mainly) on the following 
parameters: the velocity of the SB ($v$), the direction
of its trajectory ($\cos\theta$),
the SB diameter ($D_{SB}$), and the value of the 
fundamental parameter $\epsilon$. 
[Of course, while the parameters $v$, $\cos\theta$, $D_{SB}$,
can all have different values, depending on each
event, $\epsilon$ can only have one value, which
is fixed by nature, like the electron electric charge].
In terms of the parameter $\omega$ [Recall that 
$\omega \equiv \stackrel{\sim}{D}_{SB}(\theta=0)/z \sim
(\epsilon/2\times
10^{-8})^2(30\ {\rm km/s}/v)^4$], we have:
\begin{eqnarray}
Strong \ coupling \ regime: 
\ \ &\omega & \stackrel{>}{\sim}  {5 \ {\rm metres} \over D_{SB}}
\nonumber \\
&D_{SB}& \stackrel{<}{\sim} {5 \ {\rm metres}\over \cos\theta}\ \ \
Loses \ velocity\ (v_f \sim 0.5 \ {\rm km/s}) 
\nonumber \\
&D_{SB}& \stackrel{>}{\sim} {5 \ {\rm metres}\over \cos\theta}\ \ \
Retains \ velocity\ (v_f \sim v_i \ {\rm km/s}) 
\nonumber \\
\nonumber \\
Weak \ coupling \ regime: 
\ \ &\omega & \stackrel{<}{\sim} {5 \ {\rm metres} \over D_{SB}}
\nonumber \\
&\omega & \stackrel{>}{\sim} \cos\theta\ \ \ Loses \ velocity\
(v_f \sim 0.5 \ {\rm km/s})
\nonumber \\
&\omega & \stackrel{<}{\sim} \cos\theta\ \ \ Retains \ velocity\
(v_f \sim v_i \ {\rm km/s})
\end{eqnarray}
Finally, observe that the experimental limit $\epsilon \stackrel{<}{\sim}
10^{-6}$ and the lower limit on the SB initial velocity $v \stackrel{>}{\sim} 11\ {\rm km/s}$
together imply that $\omega \stackrel{<}{\sim} 10^5$. 
The main results of this section are summarized in Figure 3.
\vskip 0.7cm
\centerline{\epsfig{file=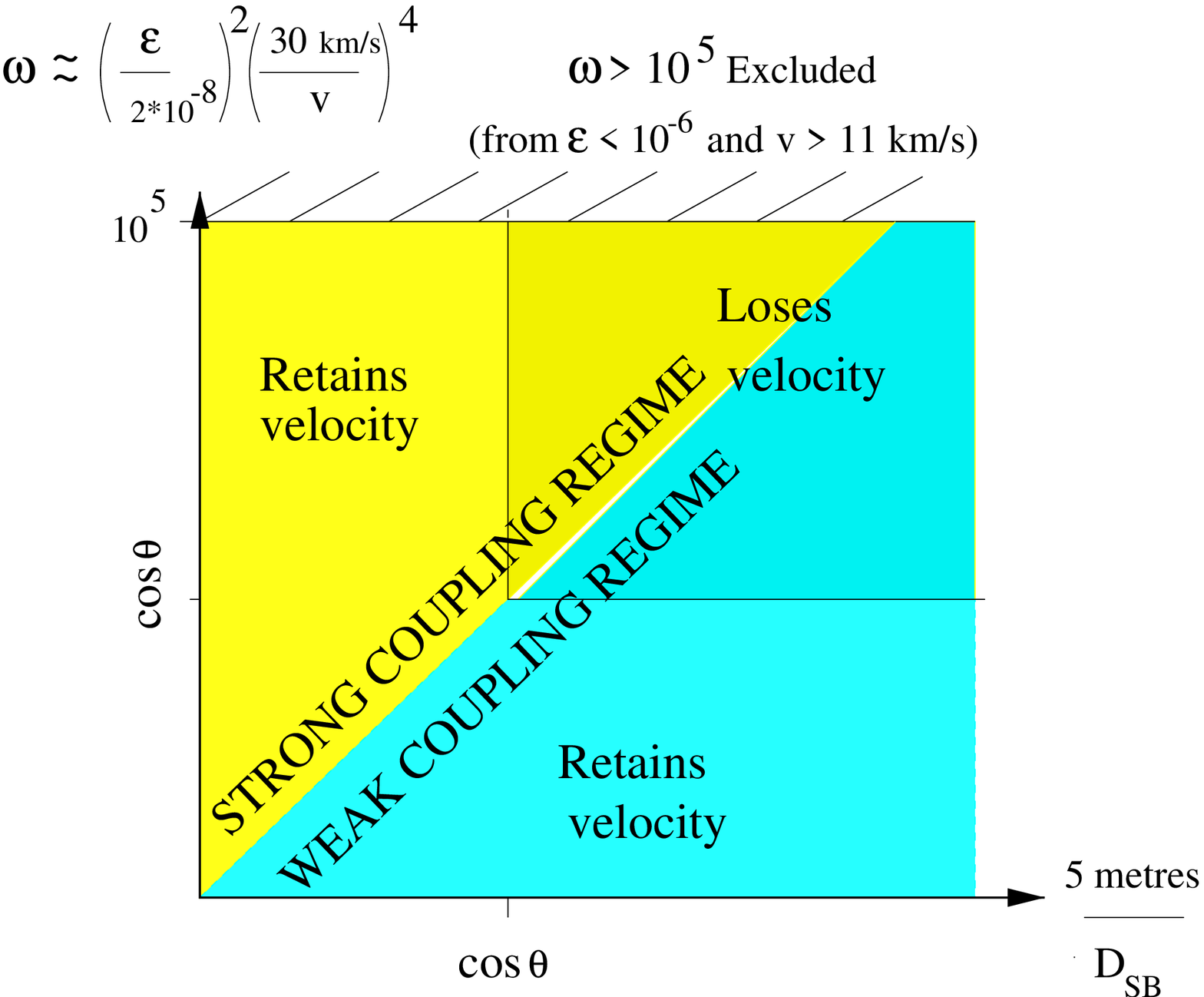,height=7cm,width=8cm}}
\vskip 0.5cm
{\small Figure 3: Pictorial summary of the results of this section.
}

\section{Observations of anomalous impact events}

There are many reported examples of atmospheric
phenomena resembling fireballs, which cannot be
due to the penetration of an ordinary meteoroid into the
atmosphere (for a review of bolides, including discussion
of these anomalous events, see Ref.\cite{bol}).
Below we discuss several examples of this strange
class of phenomena.

\vskip 0.2cm
\noindent
(i) {\it The Spanish event -- January 18, 1994.}

\vskip 0.2cm

On the early morning of 1994 January 18, a very bright luminous
object crossed the sky of Santiago de Compostela, Spain. This
event has been investigated in detail in Ref.\cite{docobo}. The
eye witnesses observed the object to be low in altitude and
velocity ($1$ to $3$ km/s). Yet, an ordinary body penetrating
deep into the atmosphere should have been quite large and luminous
when it first entered the atmosphere at high altitudes with large
cosmic velocity (between $11$ and $70$  km/s). A large ordinary body
entering the Earth's atmosphere at these velocities always
undergoes significant ablation as the surface of the body melts
and vapourises, leading to a rapid diminishing of the bodies size
and also high luminosity as the ablated material is heated to high
temperature as it dumps its kinetic energy into the surrounding
atmosphere. Such a large luminous object would have an estimated
brightness which would supersede the brightness of the Sun,
observable at distances of at least $500 \ $km. Sound
phenomena consisting of sonic booms should also have
occurred. Remarkably neither of these two expected
phenomena were observed for this event. The authors of
Ref.\cite{docobo} concluded that the object could {\it not} be a
meteoric fireball.

In addition, within a kilometre of the projected end point of the
``object's'' trajectory a ``crater'' was later
discovered. The ``crater'' had dimensions 
$29$ m $\times 13$ m and $1.5$ m deep. At the crater site, full-grown
pine trees were thrown downhill over a nearby road. Unfortunately,
due to a faulty telephone line on the $17^{th}$ and $18^{th}$ of
January (the fireball was seen on the $18^{th}$) the seismic
sensor at the nearby geophysical observatory of Santiago de Compostela
was inoperative at the crucial time.  After a careful
investigation, the authors of Ref.\cite{docobo} concluded that the
crater was most likely associated with the fireball event, but
could not definitely exclude the possibility of a landslide.
No meteorite fragments or any other unusual material was
discovered at the crater site.

\vskip 0.2cm
\noindent
(ii) {\it The Jordan event -- April 18, 2001.}

\vskip 0.2cm

On Wednesday $18^{th}$ April
2001, more than 100 people attending a funeral
procession saw a low altitude and low velocity fireball.
In fact, the object was observed to break up into two
pieces and each piece was observed to hit the ground.
The two impact sites were later examined by members
of the Jordan Astronomical Society.
The impact sites showed evidence of energy release
(broken tree, half burnt tree, sheared rocks and
burnt ground) 
\noindent
but no ordinary crater.  [This may have been
due, in part, to the hardness of the ground 
at the impact
sites].  No meteorite fragments were recovered despite
the highly localized nature of the impact sites and low
velocity of impact. For more of the
remarkable pictures and more details, see the Jordan
Astronomical Society's report\cite{jas}.
As with the 1994 Spanish event (i),
the body was apparently not observed by anyone when it 
was at high altitudes where it should have been very bright.
Overall, this
event seems to be broadly similar to the 1994 spanish
event (i). For the same reasons discussed
in (i) (above) it could not be due to an ordinary meteoric fireball.
Of course the energy release evident at the Jordan site is
another compelling reason against an ordinary meteorite interpretation.

There are many other reports of such anomalous events. For an anomalous
meteorite catalogue see Ref.\cite{andrei}.

\vskip 0.6cm

Can these anomalous meteorite events be due to
the penetration into the atmosphere of a mirror SB?
Clearly, the low impact velocity suggests
that the body has lost its cosmic velocity in
the atmosphere.
From our analysis in the previous section, we see
that a small ($D_{SB} \stackrel{<}{\sim} 5$ metres) 
mirror SB will lose its initial velocity in
the atmosphere provided that $\omega \stackrel{>}{\sim} \cos\theta$.
This suggests a rough upper bound on $\epsilon$ of
about $\epsilon \stackrel{>}{\sim}
10^{-9}$, which is certainly possible
since the current experimental limit is $\epsilon \stackrel{<}{\sim}
10^{-6}$ from orthopositronium experiments.
Furthermore, the two most puzzling features
of the small impact events
(dim at high altitudes and lack of fragments)
seem to have a natural
explanation within the mirror matter interpretation as
we will now discuss.

Ablation occurs quite rapidly
for an ordinary body because the surface quickly heats up and 
melts. The surface breaks off, quickly vaporizing as the 
surface atoms rapidly decelerate from $v_i \sim 30$ km/s
to $\sim 0$ thereby heating the surrounding atmosphere in the process.
This ablation process -- which dumps energy into 
the atmosphere -- is the origin of the light emitted when an ordinary SB enters
the Earth's atmosphere. An ordinary body only undergoes
ablation while its velocity is large enough (typically
$v \stackrel{>}{\sim} 5$ km/s) which means that
a small body ($D_{SB} \stackrel{<}{\sim} 5$ metres) can
only be bright at high altitudes (typically $h \stackrel{>}{\sim}
20$ km). At low altitudes a small ordinary body is very dim, 
and its surface has already 
cooled by the time it reaches the Earth's surface.
For this reason observations of meteorite falls are quite
rare. To summarize, an ordinary small body starts off very bright
and becomes dimmer which is the exact {\it opposite}
of what was observed for the anomalous meteorite events. 

In the case of a mirror matter SB, things are quite different.
Because the air molecules penetrate the SB, the energy of
the impacting air molecules on the SB
is distributed throughout the whole volume of the SB, not
just at its surface (much like the
distinction between a conventional oven and microwave oven!). 
A mirror body would therefore heat up {\it internally}
from the interactions of the atmospheric atoms which would
penetrate within the space-body.
Thus, initially, the rate
of ablation of a mirror matter SB can be very low, which
means that the mirror SB 
can be quite dim at high altitudes where it first enters the
atmosphere. Evidently, the non-observation of 
the Spanish and Jordan events at large distances
($\stackrel{>}{\sim} 50$ km) is explained if they
are caused by a small mirror matter SB.
Furthermore, because the mirror SB heats up internally, 
it can act as a heat reservoir. If there happens to be 
any ordinary fragments embedded within the SB,
such fragments will be heated to high temperature 
and may thereby be a source of light. This may
allow the body to appear bright to nearby observers, 
perhaps consistent with the observations.
Of course, one must check that the
mirror SB does not heat up enough so that
the whole body melts. In Ref.\cite{fy}, the relevant
calculations were done and it was shown that small
mirror SB can remain intact if it is made of highly non-volatile
material and have initial velocities near the minimum allowed
value $\sim 11$ km/s. For volatile material such as ices and/or
impacts at higher velocities, mirror SB typically heats up so much
that the whole body melts, leading to a rapid release of the SB
kinetic energy into the atmosphere -- an air burst.

The lack of ordinary fragments is also explained 
if the mirror SB is made of (or predominately of) mirror matter.
Yet, mirror matter fragments should be stopped in the ground
(and potentially extracted!)
because the small electromagnetic force implied by the 
photon-mirror photon kinetic mixing is strong enough
to oppose the force of gravity (provided that  $\epsilon
\stackrel{>}{\sim} 10^{-10}$)\cite{fy}.
After the mirror matter 
body strikes the ground its
properties depend
importantly on the sign of $\epsilon$ (for
a detailed discussion, see Ref.\cite{fy}).
For $\epsilon > 0$, mirror atoms repel ordinary atoms (at close
range) while for $\epsilon < 0$, mirror atoms attract ordinary
atoms. Hence, for positive $\epsilon$ one might expect some fragments
of mirror matter to exist right on the ground, depending perhaps
on the composition of the ground and SB and also on the velocity of impact etc.
On the other hand, for negative
epsilon mirror matter could be completely embedded within
ordinary matter, releasing energy in the process. Perhaps this release
of energy was responsible for the burning of the ground (and tree!)
at the Jordan impact site.

Another source of energy release from an impacting mirror
body might come from the build-up of
ordinary electric charge within
the body\cite{zep}. Electric charge can accumulate due
to the effect of ionizing collisions.
Air molecules (initially) strike the SB with
velocity $v \sim 30$ km/s, which implies a
kinetic energy of: 
\begin{eqnarray}
E = {1 \over 2} M_A v^2 \approx 140\left( {v \over 30 \ {\rm
km/s}}\right)^2
\ {\rm eV}
\end{eqnarray}
Clearly, the energy of the impacting air molecules
is great enough for ionizing collision to occur.
Furthermore, ionized air molecules can be trapped within
the SB due to the effects of Rutherford scattering,
while the much faster moving electrons can escape the body,
leading to a build up of ordinary electric charge within
the mirror SB.


Thus, it seems that the small anomalous meteorite events
are only anomalous if they are due to an ordinary matter
SB. If they are instead composed of mirror matter,
then they apparently fit the observations. Clearly this
is good reason to take this possible explanation
seriously. Another reason is that there is no other
known mechanism which can explain their existence!

On larger scales, we have the famous Tunguska event\cite{val}.
The Tunguska event is the only recorded example of a 
large SB ($D_{SB} \sim 100$
metres) impact. This event is characterized by
a huge release of energy at a height of about
8 km.  Recall that an ordinary or mirror SB of size $D_{SB} \sim
100$ metres does not
lose much of its initial velocity in the atmosphere
(provided that it remains intact). Presumably the Tunguska 
space-body
rapidly disintegrated at low altitudes
causing a large increase in effective
surface area of the body allowing it to quickly dump  
its kinetic energy into the atmosphere.
Remarkably no significant fragments or chemical traces
were found at the Tunguska site. Furthermore there was
evidence for smaller subsequent explosions at even
lower altitudes and some
evidence that part of the SB continued its journey after
the initial explosion\cite{val}.
While the huge explosion may have vaporized most of
the SB material the lack of even chemical traces (such
as iridium excess in the soil), small fragments, and
the evidence for smaller subsequent explosions make
the Tunguska event quite puzzling.

If the Tunguska event is due to the penetration of a mirror
SB, then it seems to be somewhat less puzzling. In this
case the lack of ordinary fragments and chemical traces
are automatically explained. The atmospheric explosion
is expected as the body heats up {\it internally} 
and melts. Previous calculations\cite{fy}
have shown that the typical height for this to occur is
of order $5-10$ km for a mirror icy body moving at $\sim 11$ km/s. 
In this case it is quite possible for small pieces of the body to survive
especially if the body is not of homogeneous (mirror) 
chemical composition, which
can lead to secondary explosions.

\section{Other applications}
\subsection{Impact craters on asteroids}

The proportion of mirror SB 
to ordinary ones in the solar system is theoretically
uncertain. However there are some reasons to think
that mirror SB could actually dominate
over the ordinary ones. First, if comets really are the
mirror SB, then estimates\cite{lev} of the number of such
bodies in the inner solar system inferred from the rate
at which new comets are seen is large and dwarfs the visible
(ordinary body) population. Second, if the 
anomalous meteorite events (including Tunguska) are due to
mirror SB this hints at a large mirror SB population. 
Indeed, recent estimates of small visible bodies
(presumably made of ordinary matter since mirror SB
would appear to be quite dark, potentially unobservable) is
surprisingly low, giving a Tunguska-like impact rate of one per
thousand years\cite{low}. So, the existence of a Tunguska sized event
within the last 100 years does suggest that the mirror SB impact rate may
exceed the ordinary body rate.
Finally, there is one more indication of a large population
of mirror SB coming from observations of crater rates on
asteroids, as we will now explain.

Impacts of mirror SB on small ordinary bodies such as asteroids or moons
is of particular interest because
of the absence of atmospheric effects.
In this case the mirror SB will lose its energy
at or below the surface. The stopping distance ($L$)
of a mirror SB can be calculated from Eq.(\ref{stop}) with
the replacement $\rho_{SB} \to \rho_{asteroid}$ and $M_A \leftrightarrow
M_{A'}$, which is
\begin{eqnarray}
L \sim {v^4 M^2_{A'}M_A \over 160\pi \rho_{asteroid} Z^2 Z'^2 \epsilon^2 e^4}
\sim \left( {v \over 30 km/s}\right)^4 \left( {10^{-9} \over \epsilon}
\right)^2 km,
\label{yy4}
\end{eqnarray}
This formula only makes sense in the region
of parameter space where $L \stackrel{>}{\sim} D_{SB}$.
In the case where $L \stackrel{<}{\sim} D_{SB}$, it means
that the energy is essentially released at
the surface leading to a surface crater, while for
$L \gg D_{SB}$ no surface crater would be formed.
If $L \stackrel{>}{\sim} D_{SB}$ and $\epsilon \stackrel{>}{\sim}
10^{-8}$ then the SB will (typically) completely melt and breakup
before releasing all of its energy within the target
object (moon/asteroid).
Roughly, the energy will be released over a significant (cone
shaped) {\it volume}.
This will cause
heating of the target object which may also melt, but
will subsequently reform.

Clearly, the impact of a mirror SB will be
quite different to an ordinary impact, even
when $L \sim D_{SB}$. This is because the energy is released
not as quickly as an ordinary impact, which may well
be expected to reduce the size of the crater, if indeed one
is formed. Of course, detailed numerical simulation work needs to be
done. Nevertheless, qualitatively, we can at least
observe that
it might be possible to infer the size of $\epsilon$
by looking for a characteristic crater size below
which there is a significant reduction in crater rates.
The point is that 
small mirror SB with sizes $D_{SB}/L \ll 1$ will
not form impact craters because the energy is released too slowly
and over too large a volume.
Of course, to have an observable effect assumes that
there is a significant proportion of mirror SB  
in the solar system, which is certainly quite possible
(as discussed above).

Taking a $v_{max}$ of $\sim 30$ km/s, then
the condition that $L \sim D_{SB}^{crit}$ implies that:
\begin{eqnarray}
\epsilon \sim 10^{-8} {\left( {v \over 30 {\rm km/s}} \right)^2
\over \sqrt{D_{SB}^{crit}/10 {\rm metres}}}
\label{p5}
\end{eqnarray}
This equation relates the value of $\epsilon$ to the
minimum size of the SB ($D_{SB}^{crit}$)
for which a crater can form.
We now need to 
relate the crater size ($D_{crater}$) to the size 
of the impacting SB ($D_{SB}$).
[Roughly, we would expect $D_{crater} \sim (10-100)D_{SB}$].
We also need to see if there is any evidence of a 
crater hiatus at a critical crater size.

Actually, there is interesting evidence 
for such a crater hiatus coming from observations
of the NEAR-Shoemaker spacecraft.
This spacecraft studied the asteroid 433 Eros
in great detail, orbiting it and eventually landing
on its surface in February 2001\footnote{
Observations on asteroids are particularly `clean'
because of the absence of secondary impacts
(i.e. impacts caused by ejected material falling
back on the body after the primary impact)
which inevitably occur for impacts on large bodies
such as the moon.}.
Analysis of the size of craters on Eros' surface
shows\cite{near1} a sharp {\it decrease} in the crater
rate for craters less than about 70 metres in diameter.
The number of craters below 30 metres in diameter is 
several orders of magnitude 
less than expected\cite{near1}.
Interpreting this reduction as the onset
of the predicted crater hiatus for mirror SB impacts
suggests an $\epsilon$ of order $10^{-7}-10^{-9}$. [A more
precise estimate would require a detailed simulation
of mirror SB impacts to determine a) more precisely $D_{SB}^{crit}$ [i.e.
improve on our rough estimate, Eq.(\ref{p5})] and
b) more precisely how $D_{crater}$ is related to $D_{SB}^{crit}$].

Taking a closer look at Eros, we can find one more tantalizing
hint supporting the mirror SB hypothesis. 
Crater `ponds' were unexpectedly observed on Eros\cite{near2,near1}.
These are flat surfaces at the bottoms of craters (for
a picture of one of these, see Ref.\cite{near2,pict}).
In fact, they are found to be extremely flat (relative
to local gravity), as if they were formed by a fluid-like motion. 
Yet, the `ponds' are currently no longer
fluid-like, since their surfaces have steep-walled grooves,
small craters and support boulders\cite{near2}. In other words,
the material comprising the `ponds' was fluid-like when
the ponds were formed, but then became solid-like.
These puzzling observations may have a simple
explanation within the mirror SB hypothesis.
The impact of a large mirror SB, large enough to
form a crater, would melt not
only the mirror SB, but also a part of the asteroid. 
The melted rock of the asteroid and/or the remnants of 
the mirror SB may have left a smooth
flat surface, depending on the viscosity of the material,
rate of cooling etc.
Indeed,
the pictures of the `ponds'\cite{near2,pict} look so much
like real ponds that it is tempting to speculate that
they are made of mirror ice, with perhaps a surface
covering of ordinary dust.  This might be possible if
the mirror SB were predominately made of mirror ices (mirror
$H_2 O$, mirror $NH_3$ etc).  After the
impact, a large part of the mirror SB vaporized, and some condensed 
remnants were left on the asteroid surface which eventually froze 
\footnote{
Of course, this explanation would only be viable if
there were some mechanism to keep the mirror matter
on the surface. Mirror matter in the solid state
can potentially exist on the surface because the
small photon-mirror photon interaction can
oppose the feeble gravity on a solid mirror fragment(c.f. Ref.\cite{fy}). 
However, while the mirror matter is in the liquid form
one might expect it to seep into the asteroid,
thereby preventing a mirror ice pond from forming on the surface.
However, it might be possible for the
mirror water to freeze at some
depth below the surface, which could then act as an impermeable barrier
to keep the mirror liquid from seeping further into
the asteroid, at least long enough for a frozen pond
to form on the surface.}.

At the present time, Eros is the only small (i.e. asteroid-sized) body which
has been mapped in detail on small scales. Clearly, we would expect 
a similar dearth of small craters to occur 
on other small bodies such
as other asteroids and the moons of Mars. Ponds could also be expected
to occur, but
they might be obscured by dust and debris, depending on the strength of
the bodies gravity.

We now turn to another indication for
mirror matter in our solar system, coming from
quite a different direction.

\subsection{Pioneer Spacecraft anomaly}

Another interesting indication for mirror matter in our
solar system comes from the Pioneer 10 and 11 spacecraft anomalies.
These spacecraft, which are identical in design, were launched in 
the early 1970's with Pioneer 10 going to Jupiter and Pioneer
11 going to Saturn.  After these planetary rendezvous, 
the two spacecraft followed
orbits to opposite ends of the solar system 
with roughly the same speed, which is now about 12 km/s.  
The trajectories of these spacecraft were carefully monitored
by a team of scientists from the Jet Propulsion
Laboratory and other institutions\cite{study}.
The dominant force on the spacecraft is, of course, the gravitational 
force, but there is also another much smaller force
coming from the solar radiation pressure -- that is, 
a force arising from the light striking the surface of the spacecraft.
However, the radiation pressure 
decreases quickly with distance from the sun, and for distances greater
than 20 AU it is low enough to allow for a 
sensitive test for anomalous forces in the solar system. The
Pioneer 11 radio system failed in 1990 when it was about 30
AU away from the Sun, while Pioneer 10 is in better shape
and is about 70 AU away from the Sun (and still transmitting!).

The Pioneer 10/11 spacecrafts are very sensitive probes
of mirror gas and dust in our solar system if 
the photon-mirror photon transition force exists\cite{fvpioneer}.
Collisions of the spacecraft with mirror particles
will lead to a drag force which will slow the spacecraft down.
This situation of an ordinary matter body (the spacecraft)
propagating through a gas of mirror particles is
a sort of `mirror image' of a mirror matter space-body
propagating through the atmosphere which was considered in
the section II.

Interestingly, careful and detailed studies\cite{study} of the motion
of Pioneer 10 and 11 have revealed
that the accelerations of {\it both} spacecrafts are anomalous 
and directed roughly towards the Sun, with magnitude,
$a_p = (8.7 \pm 1.3)\times 10^{-8}\ {\rm cm/s^2}\ $.
In other words, the spacecrafts are inexplicably slowing down!
Many explanations have been
proposed, but all have been found wanting so far.
For example, ordinary gas and dust cannot explain it
because there are rather stringent constraints on
the density of ordinary matter in our solar system
coming from its interactions with the sun's light.
However, the constraints on
mirror matter in our solar system are much weaker
because of its invisibility as far as its interactions with
ordinary light is concerned.

If the interactions of the mirror particles are strong
enough so that the mirror particles lose their relative
momentum within the spacecraft, then the drag force
on the spacecraft is proportional to the bodies cross
sectional area, Eq.(\ref{st}) [with $\rho_{air}$ replaced
by the mass density of mirror matter encountered by
the spacecraft].
The condition that the mirror particles lose their
relative momentum within the spacecraft can be
obtained from Eq.(\ref{y76}):
\begin{eqnarray}
\epsilon \stackrel{>}{\sim} 10^{-7}\left({v \over 12\
{\rm km/s}}\right)^2 \sqrt{ {{\rm centimetre}\over  D_{SB}}}
\end{eqnarray}
This is the case considered in Ref.\cite{fvpioneer}.
In this case setting $a_p = F_{drag}/M_{Pioneer}$ implies that the 
density of mirror matter in 
our solar system is about $\approx 4 \times 10^{-19}\ {\rm g/cm^3}$.
It corresponds to about 200,000 mirror hydrogen atoms 
(or equivalent) per cubic centimetre\cite{fvpioneer}. 
If the mirror gas/dust
is spherically distributed with a radius of order 100 AU, then the total
mass of mirror matter would be about that of a small planet ($\approx
10^{-6} M_{sun}$) with only about $10^{-8} M_{sun}$
within the orbit of Uranus, which is about two orders
of magnitude within present limits.  If the configuration
is disk-like rather than spherical, then the total mass of mirror
matter would obviously be even less\footnote{The requirement
that the mirror gas/dust be denser than its ordinary counterpart at 
these distances could be due
to the ordinary material having been expelled by the solar wind.}.

For values of $\epsilon$ much less than $10^{-7}$ we are in
the weak coupling regime and the drag force
is proportional to the mass of the spacecraft (assuming it 
is of uniform chemical composition).
In this case, the anomalous acceleration is given by
[using Eq.(\ref{yd}) and $a_{drag} = v dv/dx$]:
\begin{eqnarray}
a_{drag} = {Z^2 Z'^2 \epsilon^2 e^4 40\pi \rho_S \over
M_A M_{A'}^2 v^2}
\end{eqnarray}
To explain the measured anomalous acceleration requires
\begin{eqnarray}
\rho_S \sim 10^{-17} \left({3\times 10^{-9} \over \epsilon}\right)^2
\ {\rm g/cm^3}
\end{eqnarray}
Constraints on the presence of invisible matter
from the orbit of Uranus\cite{cons} suggest an upper bound on
$\rho_S$ of $\sim 10^{-17}\ {\rm g/cm^3}$ 
assuming that the invisible matter
is spherically distributed,
while, if the mirror
matter is predominately distributed on the ecliptic
(rather than spherical), then the limits on $\rho_S$
from the gravitational perturbations of the outer planets
are very much weaker.

\subsection{Earth through going events}

Hitherto we have discussed SB coming from our solar system.
For such SB, their velocity (relative
to the Earth) is limited to about $70$ km/s.
SB coming from outside the solar system would have a much larger
velocity.
A mirror SB from the galactic halo
would have a typical velocity of about 300 km/s,
while an intergalactic SB would have an even higher velocity.
Such high velocity bodies would penetrate the atmosphere without
losing significant velocity and penetrate a significant
distance underground.
Of course, such events should be very rare, but nevertheless
should occur from time to time.
Such events could cause Earthquakes, perhaps distinguishable from
ordinary Earthquakes
by the location of the epicenter (far from fault lines etc).

As well as causing Earthquakes,
it might also be possible to have a {\it through going SB},
that is, one that enters and exits the Earth.
From Eq.(\ref{yy4}), we have
\begin{eqnarray}
L \sim {v^4 M^2_{A'}M_A \over 160\pi \rho_{Earth} Z^2 Z'^2 \epsilon^2 e^4}
\sim \left( {v \over 300 km/s}\right)^4 \left( {10^{-9} \over \epsilon}
\right)^2 \ 10^4 \ km.
\end{eqnarray}
Interestingly, a recent study\cite{tep} has found evidence for two such
events (both with $L \sim 10^4$ km), which might be consistent
with such a high velocity SB if $\epsilon \sim 10^{-8} - 10^{-10}$.
Clearly, more work (and more data!) is needed.

This concludes our exploration of the solar system implications
of mirror matter.

\section{Concluding Remarks}

We have explored further the implications of mirror matter
in the solar system. These implications depend importantly
on the fundamental parameter $\epsilon$ which
describes the strength of the photon-mirror photon
interaction. We have decided to ignore interesting, but
as yet unconfirmed
hints suggesting $\epsilon \approx 10^{-6}$\cite{fg} coming from the 1990 
vacuum cavity orthopositronium experiment\cite{vac} 
and study how the physics
changes as we vary $\epsilon$ (i.e. with  $ 0 < \epsilon \stackrel{<}{\sim}
10^{-6}$). 
We have shown that observations of the anomalous 
meteorite events require $\epsilon \stackrel{>}{\sim}
10^{-9}$. In addition, a similar bound also arises
from the mirror matter explanation for the Pioneer
spacecraft anomaly (assuming that the mirror matter 
is spherically distributed).

We have argued that there are some hints that small mirror space-body actually
dominate over ordinary matter bodies in the solar system.
In fact, under this assumption,
we have found some interesting new evidence for
mirror matter in the solar system.
This arises from the observation that 
mirror space-bodies colliding with asteroids should
{\it not} leave any crater if the space-bodies are below
a certain size, the precise value depending on $\epsilon$. 
Thus, we expect a crater hiatus for craters smaller
than some characteristic size. Interestingly, such
a sharp crater reduction is in fact observed, suggesting 
that $\epsilon$ is in the range $10^{-7}-10^{-9}$.
This range is consistent with the mirror matter interpretation
of the anomalous meteorite observations (including the Tunguska and Jordan 
events) and Pioneer spacecraft anomaly.

The results of this paper have important implications for
future orthopositronium experiments 
(such as the ETH-Moscow experiment\cite{eth}).
These experiments may be able to cover much of the
interesting $\epsilon$ parameter range, especially
if $\epsilon$ happens to be larger than about $10^{-8}$.
However, if we are unlucky and $\epsilon$ happens to be below about
$10^{-8}$ then it may not be possible to reach the required sensitivity
with orthopositronium for a while.
In any case, direct detection of mirror matter is possible
at the impact sites such as the one in Jordan\cite{jas}, which
only requires that the mirror matter is stopped in
the ground (i.e. $\epsilon \stackrel{>}{\sim} 10^{-10}$\cite{fy})
and could be the best way of really proving the existence of mirror
matter. 

\vskip 1.5cm
\noindent
{\large \bf Acknowledgements}
\vskip 0.6cm
R.F wishes to thank S. Blinnikov and J. Learned 
for learned correspondence
regarding the through
going space-bodies and R. Volkas and T.L. Yoon for general
discussions.
R.F. would also like to thank G. Filewood, for some
viscous discussions.

\end{document}